\documentclass[twocolumn,showpacs,amssymb,prl]{revtex4}

\usepackage{graphicx}
\usepackage{dcolumn}
\usepackage{bm}

\def\beq{\begin{equation}}
\def\enq{\end{equation}}
\def\ba{\begin{eqnarray}}
\def\ea{\end{eqnarray}}
\def\Mesz{M\'esz\'aros~}
\def\<{<\!\!}
\def\>{\!\!>}
\def\ra{\rightarrow}

\def\vareps{\varepsilon}

\begin{document}
\input{epsf}

\title{TeV neutrinos from core collapse supernovae and hypernovae}

\author{Soebur Razzaque,$^1$ Peter M\'esz\'aros$^{1,2,3}$ and Eli
Waxman$^4$}

\affiliation{$^1$Department of Astronomy \& Astrophysics, 
Department of Physics, Pennsylvania State University, University Park,
PA 16802, USA \\ $^2$Institute for Advanced Study, Princeton, NJ
08540, USA\\ $^3$Department of Atomic Physics, E\"otv\"os University,
Hungarian Academy of Sciences, Budapest, Hungary\\ $^4$Physics
Faculty, Weizmann Institute of Science, Rehovot 76100, Israel}

\begin{abstract}
A fraction of core collapse supernovae of type Ib/c are associated
with Gamma-ray bursts, which are thought to produce highly
relativistic jets. Recently, it has been hypothesized that a larger
fraction of core collapse supernovae produce slower jets, which may
contribute to the disruption and ejection of the supernova envelope,
and explain the unusually energetic hypernovae.  We explore the TeV
neutrino signatures expected from such slower jets, and calculate the
expected detection rates with upcoming Gigaton Cherenkov
experiments. We conclude that individual jetted SNe may be detectable
from nearby galaxies.
\end{abstract}

\pacs{97.60.Bw, 96.40.Tv, 98.70.Sa}
\date{\today}
\maketitle

Core collapse supernovae (SNe) are known sources of 1-10 MeV
neutrinos, as detected from SN 1987A \cite{sn1987a}.  Previously, 100
TeV neutrinos were predicted from protons accelerated in internal
shocks in relativistic jets associated with gamma-ray burst (GRB)
event \cite{wb97}, independently of any SN connection.  Long duration
($\gtrsim$ several seconds) GRBs are now known, in several cases, to
be associated with SN type Ib/c events triggered by a stellar core
collapse which is contemporaneous with the GRB event \cite{zm04}. The
relativistic jets from such core collapses are expected to produce a
$\sim$ TeV neutrino precursor burst, while the jet is making its way
out of the collapsing stellar progenitor \cite{choked-jet}, and this
may be present even if the jet does not manage to burrow through the
stellar envelope, i.e. choked bursts which are $\nu$ bright, but
$\gamma$ dark.  These previous TeV neutrino predictions assumed jets,
with bulk Lorentz factors $\Gamma_{b} \gtrsim 100$.  The frequency of
highly relativistic jets in SN like events is at most comparable to
the ratio of GRB to the average core-collapse SN events, $\lesssim
10^{-3}$.

The deposition of bulk kinetic energy from a jet into the stellar
envelope may be a powerful contributor to the triggering of a core
collapse SN or SN Ib/c \cite{khokhlov99} and/or to the disruption of
the envelope \cite{macfadyen01}.  Only a small fraction of
core-collapse SNe appear to be associated with GRBs
\cite{berger03}. However, a significantly larger fraction of SNe may
be endowed with slower jets, of $\Gamma_{b} \sim$ few
\cite{wl01,macfadyen01,rruiz02,granot03}, which may also give rise to
``orphan'' radio afterglows not associated with detected $\gamma$-ray
emission \cite{grr04}. In this Letter, we point out that SNe endowed
with slow jets will be detectable sources of multi-GeV to TeV
neutrinos. Since the occurrence rate per galaxy of SNe is much larger
than for GRBs, the chances of one occurring nearby is significantly
enhanced. Hyperenergetic SNe (hypernovae) will be fewer in number
($\sim 10 \%$ of type Ib/c rate \cite{hypernovae}), but will have a
higher neutrino flux.  We calculate the neutrino fluence from such
slow-jet SNe within the nearest 20 Mpc, and the corresponding neutrino
detectability with upcoming kilometer scale Cherenkov detectors in
Antarctica and in the Mediterranian \cite{expts}.

{\em Jet dynamics.---} We take a jet, with $\Gamma_{b} = 10^{0.5}
\Gamma_{b,0.5}$, inside the SN progenitor star.  The corresponding jet
opening angle is $\theta_{j} \sim 1/\Gamma_{b}$.  With a variability
time $t_{v} = 0.1 t_{v,-1}$ s at the jet's base, internal shocks occur
in the jet at a collision radius $r_{j} = 2 \Gamma_{b}^{2} c t_{v}
\approx 10^{10.8} \Gamma_{b,0.5}^2 t_{v,-1}$ cm, which is inside the 
progenitor star (typically a Wolf-Rayet star of $\sim 10^{11}$ cm
radius for a SN Ib, or larger for other type II SN).  Typical jet
energy, inferred from GRBs, is $E_{j} \sim 10^{51.5} E_{51.5}$ ergs
while inside the progenitor star.  By analogy with the GRB case, we
assume that a fraction $\vareps_{e} \sim 0.1 \vareps_{e,-1}$ of
$E_{j}$ is converted into relativistic electron kinetic energy in the
internal shocks, which then synchrotron radiate in the presence of
magnetic fields representing a fraction $\vareps_{b} \sim 0.1
\vareps_{b,-1}$ of $E_{j}$.

The volume number density of leptons and baryons present in the jet is
$n'_{e} = n'_{p} = E_{j}/ (2 \pi r_{j}^2 m_{p} c^3 t_{j}) \approx
10^{20.5} E_{51.5} \Gamma_{b,0.5}^{-4} t_{j,1}^{-1} t_{v,-1}^{-2}$
cm$^{-3}$ in the comoving frame of the jet.  Here $t_{j} \approx 10
t_{j,1}$ s is the typical jet duration.  Electrons and protons are
expected to be accelerated to high energy in the internal shocks.
Typically all electrons cool down by synchrotron radiation within the
dynamic time scale $t'_{\rm dyn} = t_{v}/\Gamma_{b}$.  However, due to
the large Thomson optical depth ($\tau_{\rm Th} \sim 10^{6.6} E_{51.5}
\Gamma_{b,0.5}^{-3} t_{j,1}^{-1} t_{v,-1}^{-1}$), synchrotron photons
in the jet thermalize to a black body temperature of $E'_{\gamma} =
[15 (\hbar c)^{3} E_{j} \vareps_{e}/ (2\pi^{4} r_{j}^2 c t_{j})]^{1/4}
\approx 4.3 E_{51.5}^{1/4} \vareps_{e,-1}^{1/4}
\Gamma_{b,0.5}^{-1} t_{j,1}^{-1/4} t_{v,-1}^{-1/2}$ keV.  Also photons
from the shocked stellar plasma do not diffuse into the jet as a
result of the high $\tau_{\rm Th}$.  The volume number density of
thermalized photons produced by the shocks is $n'_{\gamma} = E_{j}
\vareps_{e}/(2 \pi r_{j}^{2} c E'_{\gamma} t_{j}) \approx 10^{24.8}
E_{51.5}^{3/4} \vareps_{e,-1}^{3/4} \Gamma_{b,0.5}^{-3} t_{j,1}^{-3/4}
t_{v,-1}^{-3/2}$ cm$^{-3}$.

The magnetic field strength in the jet is given by $B' = [4 E_{j}
\vareps_{b}/ (r_{\rm jet}^2 c t_{j})]^{1/2} \approx 10^{8.5}
E_{51.5}^{1/2} \vareps_{b,-1}^{1/2} \Gamma_{b,0.5}^{-2} t_{j,1}^{-1/2}
t_{v,-1}^{-1}$ G in the comoving frame.  The corresponding shock
acceleration time for protons is $t'_{\rm acc} = A E'_{p}/e B'
\approx 10^{-12} (E'_{p}/{\rm GeV}) A_{1} E_{51.5}^{-1/2}
\vareps_{b,-1}^{-1/2} \Gamma_{b,0.5}^{2} t_{j,1}^{1/2} t_{v,-1}$ s.  
The maximum proton energy is limited by requiring this time not
to exceed $t'_{\rm dyn}$ or other possible proton cooling process time
scales.

{\em Proton cooling time scales.---} The Bethe-Heitler (BH)
interaction $p\gamma \ra p e^{\pm}$ has a logarithmically rising cross
section $\sigma_{\rm BH} = \alpha r_{e}^{2} \{(28/9) {\rm ln}[2E'_{p}
E'_{\gamma}/ (m_{p} m_{e} c^{4})]-106/9 \}$.  The $e^{\pm}$ pairs are
produced at rest in the center of mass (c.m.) frame of the $p\gamma$
collision.  The energy lost by the proton in each interaction is given
by $\Delta E'_{p} = 2 m_{e} c^{2} \gamma'_{\rm c.m.}$, where
$\gamma'_{\rm c.m.} = (E'_{p} + E'_{\gamma})/ (m_{p}^{2} c^{4} + 2
E'_{p} E'_{\gamma})^{1/2}$ is the Lorentz boost factor of the c.m. in
the comoving frame.  The energy loss rate of the proton is given by
$dE'_p/dt'_{\rm BH} = n'_{\gamma} c \sigma_{\rm BH} \Delta
E'_{p}$. The corresponding proton cooling time is $t'_{\rm BH} =
E'_{p} / (dE'_{p}/dt') = E'_{p} /(2 n'_{\gamma} c
\sigma_{\rm BH} m_{e} c^{2} \gamma'_{\rm c.m.})$ in the comoving frame.

Protons, in the presence of the same magnetic field which is
responsible for the electron synchrotron losses, also suffer
synchrotron losses, but over a longer time scale given by $t'_{\rm
syn} = 6\pi m_{p}^{4} c^{3}/(\sigma_{\rm Th}\beta^{2} m_{e}^{2} E'_{p}
B^{'2}) \approx 3.8 (E'_{p}/{\rm GeV})^{-1} E_{51.5}^{-1}
\vareps_{b,-1}^{-1} \Gamma_{b,0.5}^{4} t_{j,1} t_{v,-1}^{2}$ s.

Protons can also transfer energy to photons by inverse Compton (IC)
scattering, because of the significantly high photon density
($n'_{\gamma}$) in the shocks.  The IC cooling time in the Thomson
limit is given by $t'_{\rm IC, Th} = 3 m_{p}^{4} c^{3}/(4 \sigma_{\rm
Th} m_{e}^{2} E'_{p} E'_{\gamma} n'_{\gamma}) \approx 3.8 (E'_{p}/{\rm
GeV})^{-1} E_{51.5}^{-1} \vareps_{e,-1}^{-1} \Gamma_{b,0.5}^{4}
t_{j,1} t_{v,-1}^{2}$ s.  In the Klein-Nishina (KN) limit, the IC
cooling time is given by $t'_{\rm IC, KN} = 3 E'_{p} E'_{\gamma}/ (4
\sigma_{\rm Th} m_{e}^{2} c^{5} n'_{\gamma}) \approx 10^{-10.5}
(E'_{p}/{\rm GeV}) E_{51.5}^{-1/2} \vareps_{e,-1}^{-1/2}
\Gamma_{b,0.5}^{2} t_{j,1}^{1/2} t_{v,-1}$ s.  The Thomson and the KN
limits apply for $E'_{p}$ much less or greater than $m_{p}^{2} c^{4}/
E'_{\gamma} \approx 10^{5.3} E_{51.5}^{-1/4} \vareps_{e,-1}^{-1/4}
\Gamma_{b,0.5} t_{j,1}^{1/4} t_{v,-1}^{1/2}$ GeV, respectively, in the
comoving frame.

\begin{figure} 
\centerline{\epsfxsize=3.45in \epsfbox{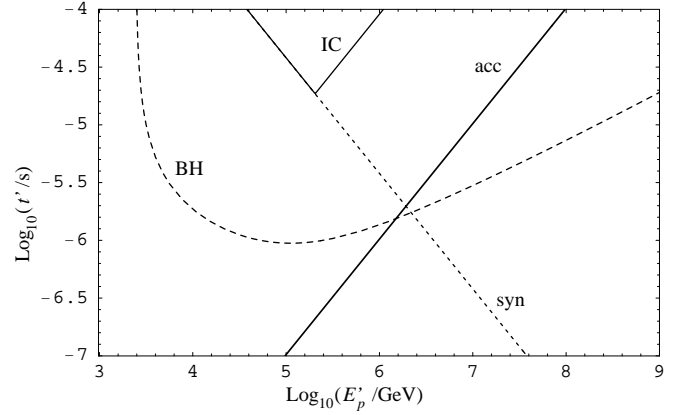}}
\caption{Proton cooling time scales for different processes, in the
comoving frame, as functions of energy.  Also shown is the proton
shock acceleration time.  Burst parameters used are $E_{\rm
sn}=10^{51.5}$ erg, $t_{j} = 10$ s, $t_{v}=10^{-1}$ s, $\Gamma_{b}=3$,
$\vareps_{e} = \vareps_{b} = 0.1$.}
\label{fig:cooltime}
\end{figure}

We have plotted the proton acceleration time and the different cooling
times in the comoving frame as functions of the proton energy in Fig.
\ref{fig:cooltime}.  Note that the BH cooling time (long dashed line)
and the synchrotron cooling time (short dashed line) are first longer
and then shorter than the maximum proton acceleration time (thick
solid line).  The corresponding maximum proton energy can be roughly
estimated, by equating the $t'_{\rm syn}$ to $t'_{\rm acc}$, as
$E'_{p,{\rm max}} \approx 10^{6.3} A_{1}^{-1} E_{51.5}^{-1/4}
\vareps_{b,-1}^{-1/4} \Gamma_{b,0.5} t_{j,1}^{1/4} t_{v,-1}^{1/2}$
GeV.

{\em Pion and muon production and cooling.---} High-energy protons
interact with thermal photons in the shocks to produce neutrinos
dominantly through $p \gamma \ra \Delta \ra n \pi^{+} \ra \mu^{+}
\nu_{\mu} \ra e^{+}~\nu_e ~{\bar \nu}_{\mu} ~\nu_{\mu}$ channel.  The
threshold proton energy at $\Delta$ is $E'_{p, {\rm th}} =
0.3/E'_{\gamma} \approx 10^{4.8} E_{51.5}^{-1/4} \vareps_{e,-1}^{-1/4}
\Gamma_{b,0.5} t_{j,1}^{1/4} t_{v,-1}^{1/2}$ GeV and the corresponding
optical depth is $\tau_{p\gamma} \approx 10^{7.8} E_{51.5}^{3/4}
\vareps_{e,-1}^{3/4} \Gamma_{b,0.5}^{-2} t_{j,1}^{-3/4}
t_{v,-1}^{-1/2}$, both evaluated in the comoving frame.  However, low
energy protons may also produce $\Delta$ by interacting with
photons from high-energy tail of the black body photon distribution
until $\tau_{p\gamma}$ falls below unity or below the optical depth
for other processes such as proton-proton ($pp$) interaction.
Similarly, protons with energy above $E'_{p, {\rm th}}$ interact with
low energy photons to produce $\Delta$.  $pp$
interactions are effective at lower energy compared to $p\gamma$
interactions. The total pion multiplicity from $pp$ interactions is of
the order unity in the energy range we are considering here
\cite{rmw03a}.  For simplicity, we assume that each $pp$ interaction
produces single pion, the same as $p\gamma$ interactions, down to the
proton energy for which the corresponding neutrino energy is above
detection threshold ($E_{\nu,{\rm th}}$) on Earth.  The available
maximum proton energy to produce $\Delta$, through $p\gamma$
interactions, is $E'_{p, {\rm max}} \approx 10^{6.3}$ GeV.

High-energy pions and muons produced in $p\gamma$ interactions do not
all decay to neutrinos, as synchrotron radiation and IC scattering
reduce the energy.  The synchrotron break energies (below which pions
and muons decay before undergoing significant cooling) are found by
equating their synchrotron cooling times ($t'_{\pi,\mu; {\rm syn}}$)
to their decay times ($\gamma'_{\pi, \mu} t_{\rm dec}$) in the
comoving frame, which are $E'_{\pi, {\rm sb}}=10^{2} E_{51.5}^{-1/2}
\vareps_{b,-1}^{-1/2} \Gamma_{b,0.5}^{2} t_{j,1}^{1/2} t_{v,-1}$ GeV
and $E'_{\mu, {\rm sb}} = 10^{0.7} E_{51.5}^{-1/2}
\vareps_{b,-1}^{-1/2} \Gamma_{b,0.5}^{2} t_{j,1}^{1/2} t_{v,-1}$ GeV,
respectively. The IC cooling time ($t'_{\rm IC}$) is comparable
to the synchrotron cooling time and the two can be combined into a
single loss rate $t_{\rm cool}^{'-1} \sim t_{\rm syn}^{'-1} + t_{\rm
IC}^{'-1}$. For IC losses in the Thomson limit $t'_{\rm syn} \approx 
t'_{\rm IC}$ while in the KN limit $t'_{\rm syn} \ll t'_{\rm IC}$, 
and the neutrino fluence in the presence of these energy losses is 
suppressed, compared to the fluence obtained assuming pions and muons
decay without any energy loss, by a factor 
\ba \zeta (E_{\pi,\mu}) &=& t'_{\pi,\mu; \rm cool} 
/t'_{\pi,\mu; \rm decay} \approx t'_{\pi,\mu; \rm cool}/t'_{\pi,\mu;
\rm syn} \nonumber \\ &=& (E_{\pi,\mu}/ E_{\pi,\mu; {\rm
sb}})^{-2} ; ~ E_{\pi,\mu} > E_{\pi,\mu; {\rm sb}}.
\label{IC-pion-reduce} \ea
Thus muons and their decay neutrinos are severely suppressed due to
their high energy electromagnetic losses. We calculate the neutrino
fluence from pion decay next.

{\em Neutrino flux.---} The observed isotropic equivalent proton
fluence from the SN jet, denoting the quantities in the observer's
frame with subscript ``ob'', at a luminosity distance $D_{L}$, is
${\cal F}_{p,{\rm ob}} = E_{j} \Gamma_{b}^{2} (1+z)/(2\pi D_{L}^{2}
E_{p,{\rm ob}}^{2} ~{\rm ln}[E_{p, {\rm max}}/E_{p, {\rm min}}])$.
Here we have assumed a shock accelerated proton energy distribution
$\propto E_{p}^{-2}$.  The energy and time in the observers frame and
in the local rest frame (at red shift $z$) are related by $E_{p} =
E_{p,{\rm ob}} (1+z)$ and $t_{j} = t_{j,{\rm ob}}/(1+z)$. The
corresponding $\nu_{\mu}$ fluence from proton interactions is
\ba {\cal F}_{\nu,{\rm ob}} &=& \frac{1}{8} ~ \frac{E_{j} \Gamma_{b}^{2}
(1+z)^{3}}{2\pi D_{L}^{2} E_{\nu}^{2} ~{\rm ln} \left[E_{\nu,
\rm{max}}/E_{\nu, {\rm min}} \right]} \nonumber \\ && \times ~\left(
\frac{E_{\nu}}{E_{\nu, \rm{sb}}} \right)^{-2} \Theta \left( E_{\nu,
{\rm min}} \le E_{\nu} \le E_{\nu, \rm{max}} \right),~
\label{nu-flux-formula} \ea
per SN burst, assuming all shock accelerated protons convert to pions.
The factor of $1/8$ is due to the fact that in the absence of energy
losses, the neutrinos produced by pion decay carry $1/8$ of the energy
lost by protons to pion production (since charged pions and neutral
pions are produced with roughly equal probability, and since the muon
neutrino produced in a pion decay carries roughly $1/4$ of the pion
energy). The neutrino energy range is then $E_{\nu, {\rm ob, min}}$ -
$E_{\nu, {\rm ob, max}} = E_{\nu, {\rm th}}$ - $0.05
\Gamma_{b} E'_{p, {\rm max}}/(1+z) \approx 10^{2.5}$ -
$10^{5.5}/(1+z)$ GeV in the observer's frame on Earth. Here we have
assumed $E_{\nu,{\rm th}} = 10^{2.5} E_{\nu,2.5}$ GeV for a typical
ice Cherenkov detector such as IceCube. For a SN at $\sim 20$ Mpc
($D_{L} \approx 10^{25.8}$ cm, $z\sim 0$), e.g. in the Virgo cluster,
the neutrino fluence would be
\ba {\cal F}_{\nu,{\rm ob}}^{*} &\approx & 10^{-3}
\left(\frac{E_{\nu,{\rm ob}}}{100\rm GeV}\right)^{-4} 
\frac{E_{51.5}^{3/2}\vareps_{b,-1}^{1/2}}{\Gamma_{b,0.5} 
D_{25.8}^{2}t_{j,1}^{1/2} t_{v,-1}} \nonumber \\ &&{\rm
GeV^{-1}cm^{-2}}; 10^{2.5} \lesssim E_{\nu,{\rm ob}}/{\rm GeV}
\lesssim 10^{5.5}.
\label{nu-flux-numbers} \ea

To calculate the diffuse neutrino flux, we sum over fluences from all
slow-jet SNe distributed over cosmological distances in Hubble time.
The SNe rate follows closely the star formation rate (SFR) which can
be modeled as ${\dot \rho}_{*} (z) = 0.32 h_{70} {\rm exp}(3.4 z)/
[{\rm exp}(3.8 z) + 45]$ M$_{\odot}$ yr$^{-1}$ Mpc$^{-3}$ per unit
comoving volume \cite{porciani01}.  Here $H_{0} = 70 h_{70}$ km
s$^{-1}$ Mpc$^{-1}$ is the Hubble constant.  For a
Friedmann-Robertson-Walker universe, the comoving volume element is
$dV/dz = 4\pi D_{L}^2 c/(1+z) |dt/dz|$ and the relation between $z$
and the cosmic time $t$ is $(dt/dz)^{-1} = -H_{0} (1+z) \sqrt{
(1+\Omega_{m}z) (1+z)^{2} - \Omega_{\Lambda} (2z+z^{2}) }$.  Here we
have used the standard $\Lambda$CDM cosmology with $\Omega_{m} = 0.3$
and $\Omega_{\Lambda} = 0.7$.

The number of SNe per unit star forming mass ($f_{\rm sn}$) depends on
the initial mass function and the SN threshold for stellar mass ($M
\sim 8$ M$_{\odot}$). A Salpeter model $\phi (M) \propto M^{-\alpha}$
with different power-law indices can generate different values for
$f_{\rm sn}$, e.g. $\approx 0.0122$ M$_{\odot}^{-1}$ for $\alpha =
2.35$ \cite{porciani01} and $\approx 8\times 10^{-3}$ M$_{\odot}^{-1}$
for $\alpha = 1.35$ \cite{hernquist03}.  We adopt the model in
Ref. \cite{porciani01} which corresponds to the local type II SNe rate
${\dot n}_{\rm sn}(z=0) = f_{\rm sn} {\dot \rho}_{*} (z=0) \approx 1.2
\times 10^{-4} h_{70}^{3}$ yr$^{-1}$ Mpc$^{-3}$ agreeing with data
\cite{madau98}.

The distribution of SNe per unit cosmic time $t$ and solid angle
$\Omega$ covered on the sky can be written as
\ba \frac{d^2 N_{\rm sn}}{dt d\Omega} = \frac{ {\dot n}_{\rm sn} (z)}
{4\pi} \frac{dV}{dz} = \frac{ {\dot n}_{\rm sn}(z) D_{L}^{2} c}
{(1+z)^{2}} \left| \frac{dt}{dz} \right|. \label{sn-dist} \ea
We assume a fraction $\xi_{\rm sn} \lesssim 1$ of all SNe involve jets
and a fraction $1/2\Gamma_{b}^2$ of all such jets are pointing towards
us. The observed diffuse SNe neutrino flux, using Eq.
(\ref{nu-flux-formula}), is then
\ba \Phi_{\nu, {\rm ob}}^{\rm diff} &=& \frac{\xi_{\rm
sn}}{2\Gamma_{b}^2} \int_{0}^{\infty} dz ~ \frac{d^2 N_{\rm sn}}{dt
d\Omega} ~{\cal F}_{\nu,{\rm ob}} (E_{\nu,{\rm ob}}) \nonumber \\ &=&
\xi_{\rm sn} \frac{c}{32\pi}
\frac{E_{j}} {E_{\nu,{\rm ob}}^{2}} \int_{0}^{\infty}
dz ~\frac{{\dot n}_{\rm sn}(z)/(1+z)} {{\rm ln} \left[ E_{\nu,
\rm{ob,max}}/E_{\nu, {\rm th}} \right]} \nonumber \\ && \times ~\left|
\frac{dt}{dz} \right| \left( \frac{E_{\nu, {\rm
ob}}[1+z]}{E_{\nu, \rm{sb}}} \right)^{-2} \nonumber \\ && \times
~\Theta \left( E_{\nu, {\rm th}} \le E_{\nu,{\rm ob}} \le E_{\nu,
\rm{ob,max}} \right).  \label{diff-flux} \ea

\begin{figure} [t]
\centerline{\epsfxsize=3.45in \epsfbox{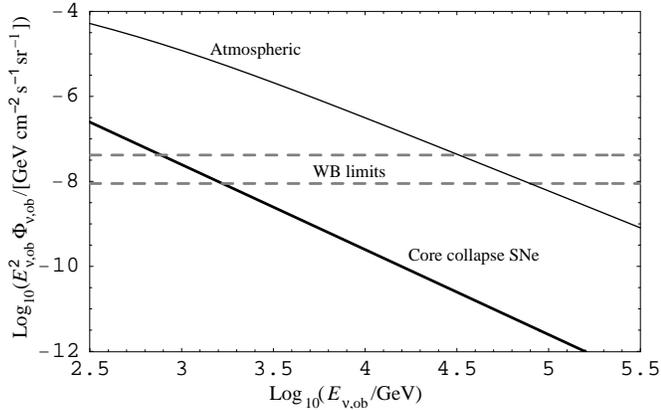}}
\caption{Diffuse muon neutrino flux on Earth from all ($\xi_{\rm sn} =
1$, $E_j=10^{51.5}$ erg) core collapse SNe (heavy solid curve).  The
dashed curves are the cosmic ray (WB) limits, and the light full curve
is the atmospheric $\nu_{\mu}$ and ${\bar \nu}_{\mu}$ flux. This shows
that the diffuse SNe flux is unlikely to be detectable with km$^2$
detectors. However, individual nearby SNe may be detectable, as
discussed below.  }
\label{fig:nuflux}
\end{figure}

We have plotted the diffuse $\nu_{\mu}$ flux from all cosmological
slow-jet SNe in Fig. \ref{fig:nuflux} by numerically integrating Eq.
(\ref{diff-flux}), assuming the maximal fraction $\xi_{\rm sn}=1$.
Also shown are the cosmic ray bounds (WB limits) on the diffuse
neutrino flux \cite{wb97}, and the atmospheric $\nu_{\mu}$, ${\bar
\nu}_{\mu}$ flux from pion and kaon decays (conventional flux,
\cite{thunman96}), compatible with AMANDA data
\cite{ahrens02}.
\ba \Phi_{\nu, {\rm ob}}^{\rm atm} = \cases{ \frac{0.012 E_{\nu,{\rm
ob}}^{-2.74}} {1+0.002 E_{\nu,{\rm ob}}}; E_{\nu,{\rm ob}} <
10^{5.8}~{\rm GeV} \cr \frac{3.8 E_{\nu,{\rm ob}}^{-3.17}}{1+0.002
E_{\nu,{\rm ob}}}; E_{\nu,{\rm ob}} > 10^{5.8}~{\rm GeV}. }
\label{atmoflux} \ea
Figure \ref{fig:nuflux} demonstrates that the diffuse flux is unlikely
to be detectable by km-scale telescopes. However, individual nearby
SNe may be detectable, as we show below.

{\em Neutrino event rate.---} The effective area ($A_{\rm eff}$) of a
Cherenkov detector depends on the zenith angle and the energy of the
muon that is created by a charged current neutrino-nucleon ($\nu N$)
interaction in the vicinity of the detector.  IceCube can trigger on
muons of energy (which is $\sim 80\%$ of the neutrino energy) above a
few hundred GeV created within $A_{\rm eff} \sim {\rm km}^2$
\cite{expts}. We assumed $E_{\nu,{\rm th}} \sim 300$ GeV, as discussed 
before.  The angular sensitivity of the IceCube detector is
$1^{\circ}$ for muon tracks with $<140^{\circ}$ zenith angles below 1
TeV. The pointing resolution gets better at higher energy.

The likeliest prospect for detection is from individual SN in nearby
starburst galaxies, such as M82 and NGC253 ($D_{L} = 3.2$ and 2.5 Mpc
in the Northern and Southern sky, respectively), over a negligible
background, using temporal and positional coincidences with optical
detections.  The SN rate in these is $\sim 0.1$ yr$^{-1}$
\cite{herrero03}, much larger than in our own galaxy or in the
Magellanic clouds.

To calculate the number of $\nu_{\mu}$ events in a km$^2$ detector
such as IceCube, we use the fluence of Eq.(\ref{nu-flux-numbers}) from
each SN in the 300 GeV-300 TeV energy range, and the full detection
probability depending on the source's angular position. This
probability depends on the Earth's shadowing effect and the energy
dependent $\nu N$ cross section inside $A_{\rm eff}$
\cite{rmw04a}. The number of $\nu_{\mu}$ events from M82 and NGC253 is 
$281~E_{51.5}^{3/2}$ and $482~E_{51.5}^{3/2}$, respectively per
individual SN with its jet pointing towards us. The timing uncertainty
in the optical detection of SN is $\sim$ 1 day. The corresponding
atmospheric background events within $1^{\circ}$ angular resolution is
0.07 for both M82 and NGC253 in the same energy range.  At the quoted
rate, a SN from one of these galaxies would be expected within five
years.
Other nearby spirals (M31, M74, M51, M101, etc, and the Virgo cluster)
will also contribute. At the standard rate of 1 SNu= $10^{-2} {\rm
yr}^{-1}$ per $10^{10}$ blue solar luminosity for average galaxies,
the roughly 4000 galaxies known within 20 Mpc would suggest a rate
$\gtrsim~1$ SN yr$^{-1}$. IceCube may detect roughly 1/5 of them at a
level of $\gtrsim 1.5$ $\nu_{\mu}$ events/SN.

Because of oscillations, neutrinos of all three flavors ($\nu_{\mu}$,
$\nu_{e}$, $\nu_{\tau}$) should reach us in equal proportion.
However, only $\nu_{\mu}$'s are emitted from the sources under
consideration and the total number of neutrino events (including
$\nu_{e}$ and $\nu_{\tau}$) will remain the same as the $\nu_{\mu}$
events we have calculated.  The lack of good directional sensitivity
for the $\nu_{e}$ and $\nu_{\tau}$ events may prevent obtaining their
positional coincidence with the SN. (Also, the $A_{\rm eff}$ for
$\nu_{e}$ and $\nu_{\tau}$ detection are somewhat smaller than
$\nu_{\mu}$.) However the timing coincidence of $\nu_{e}$ and
$\nu_{\tau}$ events with $\nu_{\mu}$ events may still be useful to
verify the neutrino oscillations at these energies, and test their
common origin.

{\em Discussion.---} While ultra relativistic jets are thought to be
responsible for the long-duration GRB associated with a small fraction
of massive core collapse SNe, mildly relativistic jets may occur in a
much larger ($\lesssim 1$) fraction of core collapse SNe. Such slow
jets may contribute to the bounce needed for ejecting the stellar
envelope, resulting in the observed SN optical display.  Late time
($\sim$ yr) radio emission may be a signature of slow jets
\cite{grr04}.  Such jets may also be needed to explain the apparent
unusually large energies of hypernovae and the anisotropies inferred
from optical polarization measurements of core collapse SNe.
Here we propose, as an independent test of this model, the observation
of TeV neutrinos, which may be detected in the near future with
IceCube and other km$^2$ neutrino Cherenkov detectors.

We thank T. Abel, J. Bahcall, D. Cowen, A. Gal-Yam and C. Pe\~na-Garay
for helpful discussions. We are grateful to Shin'ichiro Ando and John
Beacom for pointing out some errors in the published paper.  This work
was supported by NSF AST0098416, the Domus Hungarica Scientiarum et
Artium, the Monell Foundation, ISF and Minerva grants.


\begin{thebibliography}{srt}

\def\bitm{\bibitem}

\bitm{sn1987a} K.~Hirata {\it et al.,} Phys.\ Rev.\ Lett.\ {\bf 58},
1490 (1987).

\bitm{wb97} E.~Waxman and J.N.~Bahcall, Phys.\ Rev.\ Lett.\ {\bf 78},
2292 (1997).

\bitm{zm04} B.~Zhang and P.~\Mesz, Int.\ J.\ Mod.\ Phys.\ A {\bf 19}, 
2385 (2004).

\bitm{choked-jet} P.~\Mesz and E.~Waxman, Phys.\ Rev.\ Lett.\ {\bf 87},
171102 (2001); S.~Razzaque, P.~\Mesz and E.~Waxman, Phys.\ Rev.\ D
{\bf 68}, 083001 (2003).

\bitm{khokhlov99} A.M.~Khokhlov, {\it et al.,} Astrophys.\ J.\ 
{\bf 529}, L107 (1999).

\bitm{macfadyen01} A.I.~Macfadyen, S.E.~Woosley and A.~Heger, 
Astrophys.\ J.\ {\bf 550}, 410 (2001).

\bitm{berger03} E.~Berger, {\it et al,} Astrophys.\ J.\ {\bf 599}, 
408 (2003).

\bitm{wl01} E.~Waxman and A.~Loeb, Phys.\ Rev.\ Lett.\ {\bf 87}, 
071101 (2001).

\bitm{rruiz02} E.~Ramirez-Ruiz, A.~Celotti and M.J.~Rees, Mon.\ Not.\ 
R. Astron.\ Soc.\ {\bf 337}, 1349 (2002).

\bitm{granot03} J.~Granot and A.~Loeb, Astrophys.\ J.\ {\bf 593}, 
L81 (2003).

\bitm{grr04} J.~Granot and E.~Ramirez-Ruiz, astro-ph/0403421.

\bitm{hypernovae} J.S.~Bloom {\it et al.}, Astrophys.\ J.\ {\bf 506}, 
L105 (1998); B.M.S.~Hansen, Astrophys.\ J.\ {\bf 512}, L117 (1999).

\bitm{expts} J.~Ahrens {\it et al.,} Astropart.\ Phys.\ {\bf 20},
507 (2004); E.~Aslanides {\it et al.}, astro-ph/9907432.

\bitm{rmw03a} S.~Razzaque, P.~\Mesz and E.~Waxman, Phys.\ Rev.\ Lett.\
{\bf 90}, 241103 (2003).

\bitm{porciani01} C.~Porciani and P.~Madau, Astrophys.\ J.\ {\bf 548}, 
522 (2001).

\bitm{hernquist03} L.~Hernquist and V.~Springel, Mon.\ Not.\ 
R. Astron.\ Soc.\ {\bf 341}, 1253 (2003).

\bitm{madau98} P.~Madau, M.~della Valle and N.~Panagia,  Mon.\ Not.\ 
R. Astron.\ Soc.\ {\bf 297}, 17L (1998).

\bitm{thunman96} M.~Thunman {\it et al.,} Astropart.\ Phys.\ {\bf 5},
309 (1996).

\bitm{ahrens02} J.~Ahrens, {\it et al.,} Phys.\ Rev.\ D {\bf 66},
012005 (2002).

\bitm{herrero03} A.~Alonso-Herrero {\it et al.,} Aston.\ J.\ {\bf 125}, 
1210A (2003).

\bitm{rmw04a} S.~Razzaque, P.~\Mesz and E.~Waxman, Phys.\ Rev.\ D {\bf
69}, 023001 (2004).  

\end{thebibliography}
\end{document}